\documentclass[aps,prl,twocolumn,groupedaddress,showpacs]{revtex4}
\usepackage{float}
\usepackage{dcolumn}
\usepackage{amsmath}
\usepackage{amssymb}
\input{epsf}

\ifx\pdftexversion\undefined
  \usepackage{graphicx}
\else
  \usepackage[pdftex]{graphicx}
\fi

\newcommand{\ket}[1]{\left| {#1} \right\rangle}

\newcommand{\proj}[1]{\left| {#1} \right\rangle \left\langle {#1}\right |}

\def\p{\ket{\psi}}

\def\A{\mathcal{A}}
\def\B{\mathcal{B}}

\def\Tr{{\rm Tr}}

\def\be{\begin{equation}}
\def\ee{\end{equation}}
\def\bea{\begin{eqnarray}}
\def\eea{\end{eqnarray}}

\def\>{\rangle}
\def\<{\langle}

\begin{document}

\title{Entanglement in the stabilizer formalism}

\author{David Fattal$^\dagger$, Toby S. Cubitt$^o$, Yoshihisa
    Yamamoto$^\dagger$, Sergey Bravyi$^\ddagger$,
	and Isaac L. Chuang$^*$}
\affiliation{$^\dagger$\mbox{Quantum Entanglement Project, ICORP, JST}\\
	\mbox{Ginzton Laboratory, Stanford University, Stanford CA 94305}\\
$^o$\mbox{Max Planck Institut f\"ur Quantenoptik, Hans-Kopfermann-Str. 1,
Garching, Germany}\\
$^*$\mbox{Center for Bits and Atoms \& Department of Physics}\\
\mbox{Massachusetts Institute of Technology, Cambridge, MA 02139, USA}\\
$^\ddagger$\mbox{Institute for Quantum Information, California
    Institute of Technology, Pasadena, 91125 CA, USA}} 

\date{\today}

%%%%%%%%%%%%%%%%%%%%%%%%%%%%%%%%%%%%%%%%%%%%%%%%%%%%%%%%%%%%%%%%%%%%%%%%%%%%%
\begin{abstract}
We define a multi-partite entanglement measure for stabilizer
states, which can be computed efficiently from a set of generators
of the stabilizer group. Our measure applies to qubits, qudits and
continuous variables.
\end{abstract}

%\keywords{}
\maketitle

%%%%%%%%%%%%%%%%%%%%%%%%%%%%%%%%%%%%%%%%%%%%%%%%%%%%%%%%%%%%%%%%%%%%%%%%%%%%%

%%%%%%%%%%%%%%%%%%%%%%%%%%%%%%%%%%%%%%%%%%%%%%%%%%%%%%%%%%%%%%%%%%%%%%%%%%%%%
%
% File:   intro.tex
% Date:   26-Feb-04
% Author: D. Fattal, T.Cubitt, and I. Chuang <ichuang@mit.edu>
%
% Stabilizer entanglement paper
%
%%%%%%%%%%%%%%%%%%%%%%%%%%%%%%%%%%%%%%%%%%%%%%%%%%%%%%%%%%%%%%%%%%%%%%%%%%%%%

% title?  Entanglement of bipartite and multipartite stabilizer states

Entanglement is an important, quantifiable physical resource of
fundamental interest, rich with potential applications in
cryptography\cite{Ekert91a}, computation\cite{Ekert98a}, and condensed
matter systems\cite{ref:Vidal1}.  Bipartite pure state
entanglement is the best understood\cite{Vedral98a}; such states
can be asymptotically inter-converted at an exchange rate governed
by the entropy of entanglement\cite{Wootters98a} of the original and
transformed forms\cite{Bennett96b,Bennett96c}.  A partial order on the space of
such entangled states has also been found\cite{Nielsen98e}, leading
to the discovery of ways to catalyze certain transformations using
other entangled states\cite{Jonathan99a}.

However, because quantum states are generally impossible to
describe concisely, e.g. an $n$ qubit pure state may have $O(2^n)$
complex amplitudes, even well-defined measures such as the entropy
of entanglement are hard to compute.

The lack of efficiently computable entanglement measures has also
limited our understanding of the properties of entangled quantum
states shared between more than two parties.  While the maximally
entangled two-qubit singlet state plays the role of a ``gold
standard'' for bipartite states, the three-qubit GHZ fails in this
role for tripartite states\cite{Dur00b}.  In general, it is not known
how to properly ``price'' multipartite entanglement, so
inter-conversion relations are not well
understood\cite{Bennett2000a,Dur00a}.

A wide class of interesting entangled states is the set used in
quantum error correction\cite{Gottesman97a},
cluster-model\cite{ref:clusterQC} and fault-tolerant quantum
computation\cite{Preskill98b}, and many cryptographic protocols such
as secret sharing\cite{ref:secret}. These are {\em stabilizer states}
(also sometimes called {\em graph states}), and the study of their
entanglement was introduced by Hein {\it et al.}\cite{ref:graph},
using a method of graphs.  It was discovered that multipartite
entangled states fall into a variety of equivalence classes, but the
entanglement had to be quantified using the {\em Schmidt measure}
\cite{ref:graph}, which is generally computationally intractable.

% The three-qubit GHZ state, for example, seems like a natural candidate
% for a ``gold standard'' for intercoverting tripartite entanglement,
% just as the two-qubit maximally enangled singlet state serves for
% bipartite states.  But no reversible interconversion methods are known
% for multipartite states\cite{Bennett:MREGS}.  At the heart of this
% difficulty is the fact that the Schmidt decomposition is only well
% defined for bipartite states; generalizations such as the Schmidt
% measure for multipartite states are computational intractable.

Here, we introduce a new method for computing the multipartite
entanglement of any stabilizer state (including continuous variable
stabilizer states such as coherent and squeezed states).  Our measure
of entanglement is defined for multipartite states, and is equal to
the entropy of entanglement (up to a factor of two) for bipartite
states.  It can also be computed easily, requiring only a number of
elementary operations which scales polynomially with the logarithm of
the size of the Hilbert space.

%%%%%%%%%%%%%%%%%%%%%%%%%%%%%%%%%%%%%%%%%%%%%%%%%%%%%%%%%%%%%%%%%%%%%%%%%%%%%

%%%%%%%%%%%%%%%%%%%%%%%%%%%%%%%%%%%%%%%%%%%%%%%%%%%%%%%%%%%%%%%%%%%%%%%%%%%%%
%
% File:   results.tex
% Date:   26-Feb-04
% Author: D. Fattal, T.Cubitt, and I. Chuang <ichuang@mit.edu>
%
% Stabilizer entanglement paper
%
%%%%%%%%%%%%%%%%%%%%%%%%%%%%%%%%%%%%%%%%%%%%%%%%%%%%%%%%%%%%%%%%%%%%%%%%%%%%%

This result is made possible by the existence of efficient
descriptions of stabilizer states, which require only $2n^2$ bits
to specify an $n$ qubit state $|\psi\>$.  These numbers specify
the set of operators $U$ in the {\it Pauli group} (i.e. tensor
products of the identity $I$ and pauli operators $X$, $Y$, and
$Z$) such that $U|\psi\> = |\psi\>$ (they \textit{stabilize}
$\p$). These operators form a group $S$, generated by $n$
operators, which we write as $S = \<g_1, g_2, \ldots, g_n\>$. In
terms of $S$, we may express our main result in two parts as
follows.

\begin{figure}[hbtp]
    \includegraphics[width=2.25in]{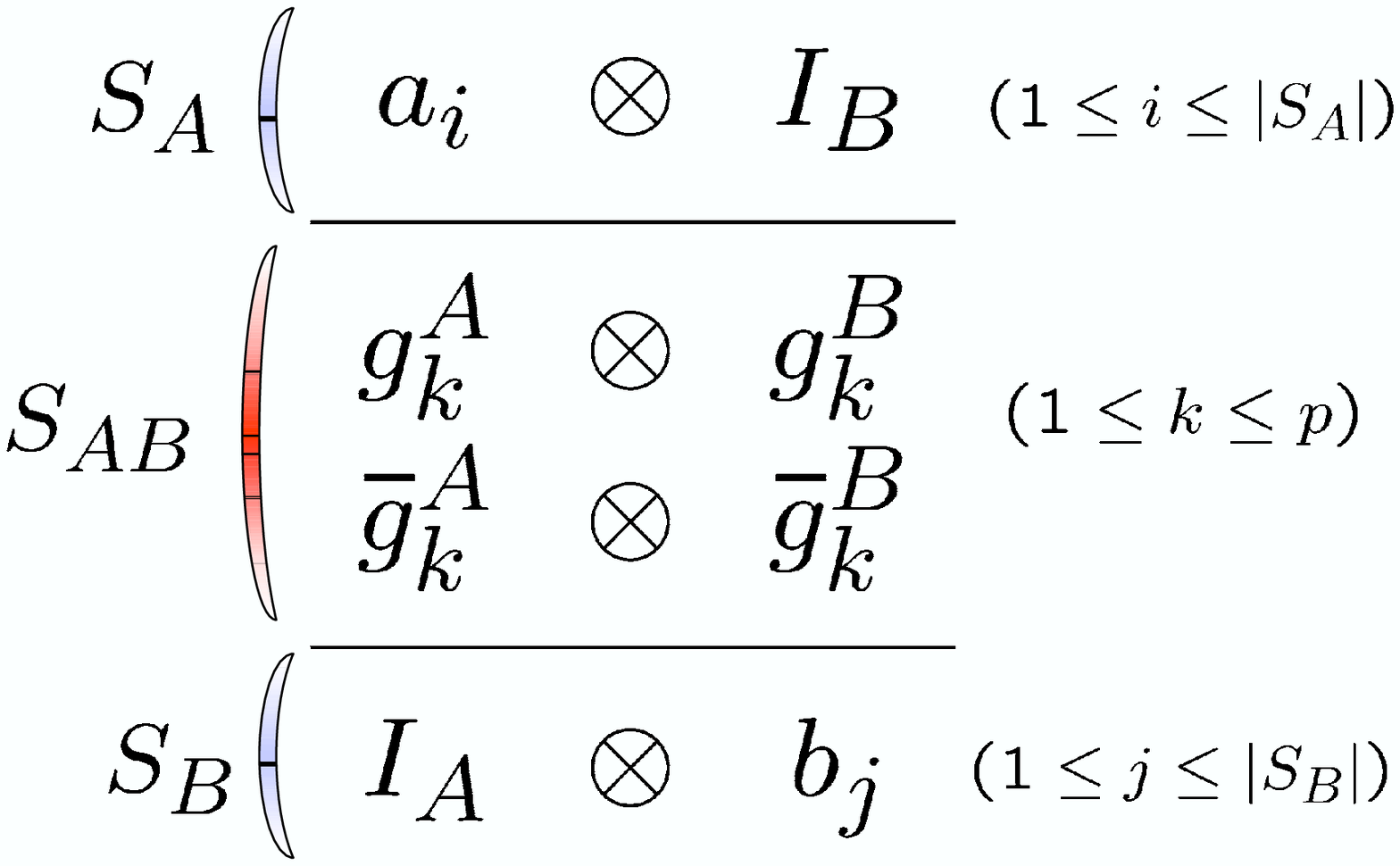}
    \caption{Canonical set of generators for a stabilizer group
    $S(\psi)$ with respect to a given partition $\{A,B\}$ of the
    qubits.  $S_A$ and $S_B$ contain the purely local information of
    $\p$. $S_{AB}$ is generated by $p$ pairs $(g_k, \bar{g}_k)$ whose
    projections on $A$ (or $B$) anticommute, but commute with all other
    generators of $S$ including elements of other pairs.}
    \label{fig:bipart}
\end{figure}

\noindent {\bf Result 1: Entanglement of Stabilizer states}: Just
as the information content of a state $|\psi\>_{AB}$ can be split
into \textit{local} information and correlations between $A$ and
$B$, the stabilizer $S$ for $|\psi\>_{AB}$ can be split into a
local subgroup $S_A \cdot S_B$ and a remaining subgroup $S_{AB}$
accounting for correlations. $S_A$ ($S_B$) correspond to
stabilizer operators that act exclusively on $A$ ($B$), as shown
in Fig.~\ref{fig:bipart}. For instance, the EPR state $|0_A\,
0_B\> + |1_A\, 1_B\>$ is stabilized by $S=\<XX, ZZ\>$, with
$S_{AB} = S$ and $S_A = S_B = \{I\}$. Furthermore, the
identification of these subgroups is simple, and require only
$O(n^3)$ steps for an $n$ qubit state.

% For example, the GHZ state $|000\> + |111\>$ has stabilizer $S = \<
% XXX, ZZI, IZZ\>$.  which splits into $S_A = S_B =

The first result of this letter is that the entropy of
entanglement $E(|\psi\>)$ is given by 
\be
    E(\p) = \frac{1}{2}|S_{AB}|
\,, 
\label{eq:first_result}
\ee 
where $|S_{AB}|$ is the rank of $S_{AB}$, meaning the size
of its minimal generating set.  For the EPR state, $|S_{AB}|=2$
which correctly gives $E=1$. For the three qubit GHZ state $|000\>
+ |111\>$, where $S = \<XXX,ZZI,IZZ\>$, with respect to partition
$A=12$ and $B=3$, we find $S_A = \<ZZI\>$, $S_B = \{I\}$, and
$S_{AB}=\<XXX,IZZ\>$, so we again obtain the correct result that
$E=1$, since $|S_{AB}|=2$.

Importantly, this expression for $E$ is easily computable; it
requires only $O(n^3)$ operations.  This is fundamentally because
stabilizer states can be efficiently described in terms of the
generators of their stabilizers.  Furthermore, this stabilizer
formalism give a constructive and efficient method to transform
any bipartite stabilizer state by local unitary operations into
$E$ independent EPR pairs.  These properties are proven below.

\noindent {\bf Result 2: Multipartite entanglement}: The
stabilizer methods also apply to characterize the multipartite
entanglement of stabilizer states.  For an $n$ qubit state
$|\psi\>$ stabilized by $S$ and split into $k$ partitions ${\cal
A} = \{A_1, A_2, \ldots, A_k\}$, we introduce a new, simple,
measure for multipartite entanglement, 
\be
    e_{\cal A}(\p) = n - \left|\prod_{j=1}^k S_j \right|
\,, 
\label{eq:def_e_multi}
\ee 
where $S_j$ contains the \textit{local} operations of $S$
that act as identity on the partition $A_j$.

$e_{\cal A}$ is a true measure of multipartite entanglement.  It
is an entanglement monotone, meaning that it does not increase
under relevant local operations and classical communications.  For
finer partitions, this entanglement measure is smaller, and it
reduces to twice the entropy of entanglement for bipartite states.

\begin{figure}[tbp]
    \includegraphics[width=3.5in]{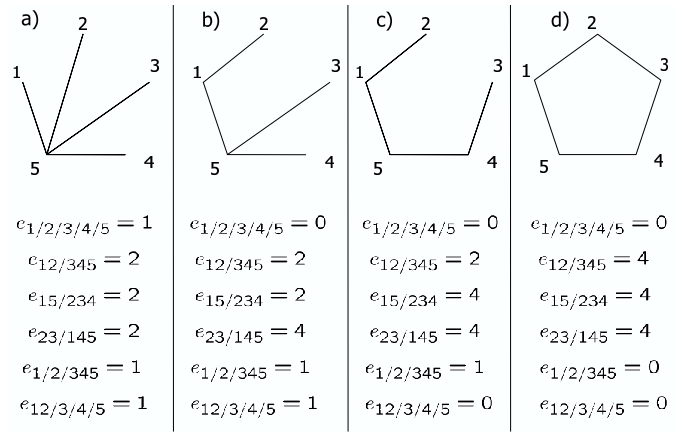}
    \caption{Application of the measure $e_{\A}$ to the classification
    of graph (stabilizer) states.  The measure is shown only for the
    relevant partitions. A complete study would have to consider all
    partitions and relabelling of the qubits. From state (a) (GHZ) to
    state (d) (cluster), entanglement becomes more "localized" and
    robust against measurement of local operators.}  \label{fig:graph}
\end{figure}

Finally in contrast to previously studied measures, it can be
computed in $O(k \cdot n^3)$ steps.  For example, Hein {\it et\,
al.} utilize the Schmidt measure to characterize the entanglement
of a class of graph states \cite{ref:graph}. The Schmidt measure
requires a difficult optimization for its computation, limiting
current studies to a small number of qubits.  Graph states are
also stabilizer states, and with this new stabilizer method, prior
graph state equivalence classes with respect to local unitaries
can be retrieved, using only simple manual computations; an
example is shown in Figure~\ref{fig:graph}.

%\input{proofs_2}
%%%%%%%%%%%%%%%%%%%%%%%%%%%%%%%%%%%%%%%%%%%%%%%%%%%%%%%%%%%%%%%%%%%%%%%%%%%%%
%
% File:   proofs.tex
% Date:   26-Feb-04
% Author: D. Fattal, T.Cubitt, and I. Chuang <ichuang@mit.edu>
%
% Stabilizer entanglement paper
%
%%%%%%%%%%%%%%%%%%%%%%%%%%%%%%%%%%%%%%%%%%%%%%%%%%%%%%%%%%%%%%%%%%%%%%%%%%%%%

We now prove the above two parts of our result.

\noindent {\bf Proof 1:} The entropy of entanglement of a bipartite state
$\p_{AB}$ is defined as:
\be
    E(\p) \equiv -\Tr\,(\rho_B \, \log \, \rho_B) 
\,, 
\ee
where $\rho_B = \Tr_A \,(\proj{\psi})$.  Since $\rho = |\psi\>\<\psi|$
is a stabilizer state, such that $g\rho = \rho$ for all $g\in S$, we
may write it as
\be
	\rho = \frac{1}{2^n} \sum_{g\in S} g
\,.
\ee
The partial trace over $A$ thus gives the reduced density matrix
\be
	\rho_B = \frac{1}{2^{n_B}} \sum_{g\in S_B} g
\,,
\ee
where $S_B$ is the subset of elements in $S$ which are nonzero when
traced over $A$.  The entropy of $\rho_B$ is thus 
\be
	E(\p) = n_B - |S_B|
\,,
\label{eq:EfromnASA}
\ee
so the entanglement now reduces to computing the rank of $S_B$.

This is accomplished most conveniently by using knowledge about the
structure of the stabilizer $S$ for $|\psi\>_{AB}$ with respect to the
partition $\{A,B\}$.  Let $P_A$ be the map that takes $g_A \otimes g_B
\in S $ onto $g_A \otimes I_B$; this a projection operator, such that 
$S_B = {\rm Ker}\, P_A$. We construct a list of generators for $S$, by
first including $|S_A|$ generators $a_i \otimes I_B$ of $S_A$ and
$|S_B|$ generators $I_A \otimes b_j$ of $S_B$. Together, these
generate the subgroup $S_{loc} \equiv S_A \cdot S_B$ that we call the
\textit{local} subgroup of $S$. Assuming the list is completed with
$e_{AB} \equiv n-|S_A|-|S_B|$ operators generating the subgroup
$S_{AB}$, then, $S$ can be decomposed as $S = S_A \cdot S_B \cdot
S_{AB}$.  Note that $|P_A(S)|=n-|S_B|$. $P_A(S)$ is in general a
non-abelian subgroup of the Pauli group.

A good choice of operators to generate $S_{AB}$ can be found by
studying the structure of subgroups such as $P_A(S)$.  First, it is
helpful to define, for an arbitrary group $G$ of Pauli operators, the
\textit{compatibility index} $c(G)$ of $G$ as the maximum rank of
an abelian subgroup of $G$. Note that $1 \leq c(G) \leq |G|$. For
later convenience, also let the \textit{incompatibility index} of $G$
be $p(G) \equiv |G|-c(G)$.  The key insight into the subgroup
structure (as illustrated by Fig.~\ref{fig:bipart}) is given by the
following theorem:

{\bf Theorem 1:} The generators for stabilizer $S$ of a bipartite
state can always be brought into the {\em canonical form}:
\be
    S = \<a_i \otimes I_B, I_A \otimes b_j,g_k,\bar{g}_k\>
\,,
\ee
where the first two subsets generate $S_A$ and $S_B$, and the last two
generate $S_{AB}$.  These generators of $S_{AB}$ collect into $p =
p(S_{AB})$ anti-commuting pairs $(g_k, \bar{g}_k)$, where $P_A(g_k)$
commute with all canonical generators of $S$ except $\bar{g}_k$, and
$P_A(\bar{g}_k)$ commute with all canonical generators of $S$ except
$g_k$.  $\Box$

This theorem implies that a stabilizer state can be transformed into
$p$ independent Bell pairs by local unitaries.  A corollary of this is
the relation $|S_A| + 2p + |S_B| = n$. Since $|S_A| + p \leq n_A$ and
$|S_B| + p \leq n_B$, it follows that
\be 
	p \,= \,n_A - |S_A| \,= \,n_B - |S_B|
%	\,=\, e_{AB}/2
	\,=\, \frac{|S_{AB}|}{2}
\,.
\label{eq:nSA}
\ee
It is also useful to know that since $|P_A(S)| = n - |S_B|$,
$p \,= \,|P_A(S)| - n_A \,= \,|P_B(S)| - n_B$.

Returing to our computation of the entanglement $E(\p)$, we now employ
Eq.(\ref{eq:nSA}) in Eq.(\ref{eq:EfromnASA}) and find that $E(\p) = p
= |S_{AB}|/2$, as claimed in Eq.(\ref{eq:first_result}).

The formalism used above shows that the problem of computing $E(\p)$
reduces to the search of anti-commuting pairs in the projections on
$A$ or $B$ of the generators of $S$. This takes $O(n^2 \cdot {\rm min}
(n_A, n_B))$ computation time and uses $2n^2$ storage
bits. Equivalently, one can compute the rank of $P_A(S)$, which is the
rank of a $n \times 2n$ matrix with elements in $\mathbb{Z}_2$
\cite{Gottesman97a}.

The quantity $E(\p)$ has a particularly simple interpretation for
graph states. In this case the group $S$ has generators
$g_j=X_j\prod_{k}' Z_k$, where the product is over all nearest
neighbors of $j$. An arbitrary element $g\in S$ can be written as
$g=\prod_{j=1}^n (g_j)^{x_j}$ for some binary vector
$x=(x_1,\ldots,x_n)$.  Element $g$ belongs to $S_A$ under certain
circumstances. First of all we must have $x_j=0$ for all $j\in
B$. Then $g$ acts as an identity operation on qubit $j\in B$ iff $j$
has even number of neighbors $k\in A$ with $x_k=1$. This requirement
is equivalent to the $\mathbb{Z}_2$-linear constraint $\sum_{k\in A}
\Gamma_{j,k}\, x_k = 0$ with $\Gamma$ being the adjacency matrix of the
graph.  Thus elements of $S_A$ are in one-to-one correspondence with
zero vectors of an adjacency submatrix $\Gamma_{B,A}$ between $B$ and
$A$.  This proves that $E(\p)$ may also be computed as the binary rank
of $\Gamma_{B,A}$, for graph states.  Bipartite entanglement in
stabilizer states can thus be visualized as arising from graph edges
crossing between the two partitions.  Note, however, that the {\em
number} of such edges does not directly give the entanglement, unless
the graph is first put into the canonical form given by Theorem~1.

% in terms of graph states, this is the rank of the
%off-diagonal part of the adjacency matrix for the partitioned graph.
%Note that the canonical form for stabilizers presented here in
%Theorem~1 is different from that used in graph states.
%\cite{Bravyi04a}.

%%%%%%%%%%%%%%%%%%%%%%%%%%%%%%%%%%%%%%%%%%%%%%%%%%%%%%%%%%%%%%%%%%%%%%%%%%%%%

We now return to prove Theorem~1, in three steps.

\textbf{Lemma 1:} $c(G) \geq \frac{|G|}{2}$ \,($p(G) \leq
\frac{|G|}{2}$)

\textit{Proof:} Denote by $C$ a maximum abelian subgroup of $G$,
generated by $c(G)$ elements $c_j$, and by $\bar{C}$ the subgroup
of $G$ generated by $|G|-c(G)$ elements $\bar{c}_k$ such that
$G=\<c_j,\bar{c}_k\>$. Up to a multiplication by elements of $C$,
each operator $\bar{c}_k$ can be made to commute with all but one
of the $c_j$. But then by the pigeon hole principle, if $c(g) <
\frac{|G|}{2}$, we can find $k_1 \neq k_2$ such that
$\bar{c}_{k_1}$ and $\bar{c}_{k_2}$ anti-commute with the same
$c_j$, so that the product $\bar{c}_{k_1} \cdot \bar{c}_{k_2}$
would commute with $c_j$ (as well as with all the other generators
of $C$), and hence $C$ would not be the maximum abelian subgroup
of $G$. $\Box$

\textbf{Lemma 2:} We can choose the generators of $G$ to be
$\{g_j\}_{1..\,c(G)} \cup \{\bar{g}_j\}_{1..\,|G|-c(G)}$, where
$g_j$ commute with all operators except $\bar{g}_j$, and
$\bar{g}_j$ commute with all operators except $g_j$.

\textit{Proof:} From Lemma~1, we know that the generators can be
organized as $c(G)$ operators $g_j$ generating $C(G)$ and $|G|-c(G)$
operators $\bar{g}_k$ generating $\bar{C}(G)$, and also that each
$\bar{g}_j$ can be made to anti-commute with $g_j$ only. We now add a
slight modification, recursively.  Suppose Lemma~1 is obeyed if we keep
only the first $m$ anti-commuting pairs $(g_k, \bar{g}_k)$. Note that
$g_m$ commute with all generators of $G$ except $\bar{g}_m$, and
$\bar{g}_m$ commutes with all $g_{k \neq m}$ and $\bar{g}_{k \leq
m}$. If $\bar{g}_{m+1}$ and $\bar{g}_m$ do not commute, we redefine
$\bar{g}_{m+1}$ to be $g_m \cdot \bar{g}_{m+1}$, so that up to this
change, the Lemma is obeyed for the first $m+1$ pairs. $\Box$

\begin{figure}[thbp]
    \includegraphics[width=2.5in]{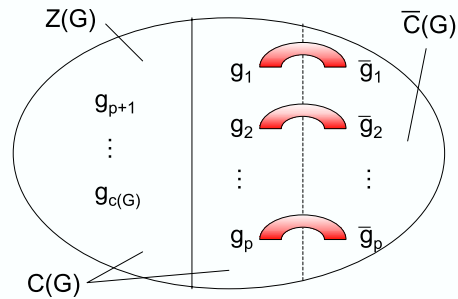}
    \caption{Structure of a subgroup G of the Pauli group. $C(G)$ is a maximum abelian subgroup of S. It
    always includes the center $Z(G)$ of G.}
    \label{fig:potato}
\end{figure}

Note that the unpaired operators $\{g_j,\, |G|-c(G)+1 \leq j \leq
c(G)\}$ generate the center of $G$, the subgroup $Z(G)$ that
commutes with all elements of $G$ (see Fig. \ref{fig:potato}).

The technical result that unravels the appropriate structure of
$S$ is the following:

\textbf{Lemma~3:} The center of $P_A(S_{AB})$ is trivial.

\textit{Proof:} Let $z$ denote the rank of the center of
$P_A(S_{AB})$, and $p$ its incompatibility index. We re-organize the
generators of $S_{AB}$ so that their projection on $A$ obeys Lemma
2. Note that their projections on $B$ have a corresponding structure,
because the generators of $S_{AB}$ commute. Then taking into account
the generators of $S_A$ and $S_B$, we find $|S_A|+z+p$ independent
commuting operators on $A$, and $|S_B|+z+p$ such operators on
$B$. Therefore the following inequalities must hold:
\bea
    & |S_A|+z+p  & \leq  n_A
\\
    & |S_B|+z+p & \leq  n_B
\\
    \Rightarrow &  |S_A|+|S_B|+2z+2p & \leq n
\,. \eea However, a simple generator count yields
$|S_A|+|S_B|+2p+z = n$, so that $z=0$. The construction used in the
proof of Lemma~3 brings $S$ into the canonical form of Theorem~1.
$\Box$

%%%%%%%%%%%%%%%%%%%%%%%%%%%%%%%%%%%%%%%%%%%%%%%%%%%%%%%%%%%%%%%%%%%%%%%%%%%%%

\noindent {\bf Proof 2:} We now turn to the more difficult
problem of finding a multi-partite entanglement measure for
stabilizer states. This will be done by a generalization of the
local subspace of S to the case of multi-partitions.

Consider a $k$-partition $\A=\{A_1,...,A_k\}$ of the $n$ qubits. We
denote the projection on $A_j$ by $P_j$ for short. We define the
subgroups $S_j$ of $S$ as $S_j \equiv \{g \in S, P_j(g)=I \}$, that is
$S_j$ is the kernel of $P_j$. We further define the local subgroup
$S_{loc}$ of $S$ as 
\be 
	S_{loc} \equiv \prod_j S_j
\,.
\ee
In the bipartite case, $S_{loc} = S_A \cdot S_B$. A qualitative
difference between the bipartite and multi-partite case is that the
subgroups $S_j$ might overlap. Therefore, it is harder to find a
canonical structure for $S$ when $k\geq 3$.  Nevertheless, the
bi-partite case can be generalized to define a $k$-partite
entanglement measure $e_{\A}$ as in Eq.(\ref{eq:def_e_multi}), that is
$e_{\A} \equiv n- |S_{loc}|$.  For a bipartition, $e_{\A}$ reduces to
$e_{AB}$ which is twice the entropy of entanglement of $\p$.

To prove that $e_{\A}$ is actually an entanglement monotone, first
note that each $S_j$ and a fortiori $S_{loc}$ are invariant under
local unitaries. Then note that the measurement of a local Pauli
operator $M$ can only increase $|S_{loc}|$. Simply, if $M$ commutes
with $S_{loc}$, then $S_{loc}$ is contained in the new local
subgroup. If not, then $M$ replaces one generator of $S_{loc}$ in
the list of generators of the post-measurement state, but since
$M$ itself is local, the new local subgroup has not decreased in
size. Finally, adding separable ancilla qubits to the system
increases $n$ and $|S_{loc}|$ by the same amount, and leaves the
difference invariant.

The entanglement measure $e_{\A}$ has also nice properties with
respect to partitions. We say that a partition $\A$ is finer than
a partition $\B$ ($\A \prec \B$) if every $A_i$ is contained in a
$B_j$.
It is easy to see that if $\A \prec \B$, then $e_{\A} \leq
e_{\B}$. Simply, every $B_j$ is a union of some $A_i$, and
therefore $S^{\B}_{loc} \subset S^{\A}_{loc}$.
Since each $S_j$ can be found in $O(n_j^3)$ computational steps,
the measure $e_{\A}$ can be computed in $O(k \cdot {\rm max}(n_j^3))$
time, requiring $2n^2$ bits of storage.

%%%%%%%%%%%%%%%%%%%%%%%%%%%%%%%%%%%%%%%%%%%%%%%%%%%%%%%%%%%%%%%%%%%%%%%%%%%%%
%
% File:   concl.tex
% Date:   26-Feb-04
% Author: D. Fattal, T.Cubitt, and I. Chuang <ichuang@mit.edu>
%
% Stabilizer entanglement paper
%
%%%%%%%%%%%%%%%%%%%%%%%%%%%%%%%%%%%%%%%%%%%%%%%%%%%%%%%%%%%%%%%%%%%%%%%%%%%%%

In summary, we have developed a mathematical formalism to efficiently
study the entanglement properties of stabilizer qubit states, which
were already known to have an efficient classical description. Among
other applications, this formalism might be useful to study
entanglement in a quantum computation involving stabilizer codes. It
could also be used to guide the construction of bipartite and
multi-partite entanglement witnesses as combinations of stabilizer
group generators. From a more fundamental point of view, it gives
some insight into the problem of understanding multi-partite
entanglement.

As a final remark, we point out that this formalism is straightforward
to generalize to qudits and continuous variable (CV) stabilizer
states. For CV stabilizer states (including gaussian states), the
Pauli group is replaced by the Heisenberg-Weyl group of displacement
operators, but the formalism is the same.

The authors gratefully acknowledge the hospitality of the Les Houches
summer school, where this work was performed at the session on quantum
information and quantum entanglement, directed by Jean-Michel Raimond
and Daniel Esteve.  This work was partially funded by the NSF Center
for Bits and Atoms Contract No. CCR-0122419.

%%%%%%%%%%%%%%%%%%%%%%%%%%%%%%%%%%%%%%%%%%%%%%%%%%%%%%%%%%%%%%%%%%%%%%%%%%%%%

\bibliography{stab}

%%%%%%%%%%%%%%%%%%%%%%%%%%%%%%%%%%%%%%%%%%%%%%%%%%%%%%%%%%%%%%%%%%%%%%%%%%%%%

\end{document}